\input{aipcheck}

\documentclass[
    ,final            
  ]
  {aipproc}

\layoutstyle{6x9}

\begin{document}

\title{Relativistic diffusion model and analysis of large transverse momentum distributions}


\author{Naomichi Suzuki}{
  address={Department of Comprehensive Management, Matsumoto University, Matsumoto 390-1295, Japan}
}

\author{Minoru Biyajima}{
  address={Department of Physics, Shinshu University, Matsumoto, 390-8621, Japan}
}


\begin{abstract}
In order to describe large transverse momentum ($p_T$) distributions observed in high energy 
nucleus-nucleus collisions, a stochastic model in the three dimensional rapidity space is  introduced.  
 The fundamental solution of the radial symmetric diffusion equation is Gaussian-like 
 in radial rapidity. 
 We can also derive a $p_T$ or radial rapidity distribution function, 
 where a distribution of emission center is taken into account.
It is applied to the analysis of observed large $p_T$ distributions of charged particles.  
It is shown that our model approaches to a power function of $p_T$ in the high transverse 
momentum limit. 
\end{abstract}

\maketitle


\section{Introduction}
In high energy nucleus-nucleus collisions, colliding energy $\sqrt{s_{NN}}$ per nucleon grows up to  $\sqrt{s_{NN}}=200$ GeV, and secondary particles with transverse momentum $p_T$ more than 10 GeV/c are observed. The observed $p_T$ distributions have long tail compared with exponential distribution in $p_T$. 

As is well known, the fundamental solution of stochastic process is Gaussian if variables in the Euclidian space are used.  Therefore, as long as we consider stochastic equations in the transverse momentum space, it is very hard to describe observed $p_T$ distributions. This fact suggests that a relativistic approach to the stochastic process would be needed.
 
In Ref.~\cite{minh73}, an empirical formula for large $p_T$ distributions at polar angle $\theta=\pi/2$, 
\begin{eqnarray}
  E\frac{d^3\sigma}{d^3p}\left|_{\theta=\pi/2} \right. &=& A\exp[-y_T^2/(2L_T)],
    \nonumber \\
  y_T &=& \frac{1}{2}\ln \frac{E+|{\bf p}_T|}{E-|{\bf p}_T|},    \label{eq.int1}
\end{eqnarray}
was proposed from the analogy of Landau's hydrodynamical model. Polar angle $\theta$ ($0\le \theta \le \pi$) is measured from the beam direction of colliding nuclei or incident particles.  In Eq.~(\ref{eq.int1}), $E$ denotes energy of an observed particle, $L_T$ is a parameter, and $y_T$ is called the "transverse rapidity".  Equation (\ref{eq.int1}) well describes  large $p_T$ distributions for $p+p\rightarrow \pi^0 + X$ and $p+p\rightarrow \pi^\pm + X$. However, it cannot be derived from the hydrodynamical model.  

As for the relativistic approach to the stochastic equation, we consider the diffusion equation 
in the three dimensional rapidity space or Lobachevsky space, which is non-Euclidean. 
In order to specify longitudinal and transverse expansions, it would be appropriate 
to consider the diffusion equation in the geodesic cylindrical coordinate,
where longitudinal rapidity $y$, transverse rapidity $\xi$ and azimuthal angle $\phi$ are used. 
The longitudinal rapidity $y$, and the transverse rapidity $\xi$ are defined respectively as,
 \begin{eqnarray*}
   y=\ln\frac{E+p_L}{m_T}, \quad \xi=\ln\frac{m_T+|{\bf p}_T|}{m},
 \end{eqnarray*}
where $E, p_L$, ${\bf p}_T$ and $m$ denote energy, longitudinal momentum, transverse momentum, and mass of the observed particle, respectively, and $m_T=\sqrt{{\bf p}_T^2 + m^2}$.
  
The diffusion equation in the geodesic cylindrical coordinate is given by,
\begin{eqnarray}
  \frac{\partial f}{\partial t} = \frac{D}{\cosh^2\!\xi}\,\frac{\partial^2 f}{\partial y^2} 
         + \frac{D}{\sinh\xi \cosh\xi}\frac{\partial}{\partial \xi} 
          \Bigl(\sinh\xi\cosh\xi\frac{\partial f}{\partial \xi}\Bigr) 
          + \frac{D}{\sinh^2\,\xi}\frac{\partial^2 f}{\partial \phi^2}, 
              \label{eq.int2}
\end{eqnarray}
where $D$ denotes a diffusion constant.  However, we cannot solve Eq.(\ref{eq.int2}) 
analytically at present. Therefore, we should consider somewhat simpler case.

  We have proposed the relativistic diffusion model, and analyzed large $p_T$ distributions for charged particles in $Au+Au$ collisions~\cite{suzu04}, where a radial flow effect is not included. 
  
  In section 2, the relativistic diffusion model is briefly explained. A distribution of the initial radial rapidity is taken into account. It would correspond to the distribution of radial flow. In section 3, large $p_T$ distributions for charged particles observed at RHIC~\cite{adam03,adle04} are analyzed.  The magnitude of radial flow is  estimated from $p_T$ distributions. In section 4, high transverse momentum limit of our model is taken and relation to Hagedorn's model for $p_T$ distribution inspired by the QCD is discussed.  Final section is devoted to summary and discussions.

\section{Diffusion equation with radial symmetry in the three dimensional rapidity space}

For simplicity, we consider the diffusion equation with radial symmetry in the geodesic polar coordinate system,
\begin{eqnarray}
  \frac{\partial f}{\partial t}= \frac{D}{\sinh^2\!\rho}\, 
      \frac{\partial}{\partial \rho}
       \Bigl( \sinh^2\!\rho\, \frac{\partial f}{\partial \rho} \Bigr),           \label{eq.dif1}
\end{eqnarray}
with an initial condition,
\begin{eqnarray}
    f(\rho,t=0)= \frac{\delta(\rho-\rho_0)}
     {4\pi\sinh^2\!\rho}.     \label{eq.dif2}
\end{eqnarray}

In Eq.~(\ref{eq.dif1}), $\rho$ denotes the radial rapidity, which is written with energy $E$, momentum ${\bf p}$ 
and mass $m$ of observed particle as,  
\begin{eqnarray}
   \rho=\ln \frac{E+|{\bf p}|}{m}.    \label{eq.dif3}
\end{eqnarray}
Inversely, energy and momentum are written respectively as
\begin{eqnarray}
    E = m\cosh\rho,   \qquad
       |{\bf p}| = \sqrt{p_L^2+ {\bf p}_T^2} = m\sinh\rho.   \label{eq.dif4}
\end{eqnarray}

The solution of  Eq.~(\ref{eq.dif1}) with the initial condition (\ref{eq.dif2}) is 
given~\cite{suzu04b} by
\begin{eqnarray}
  f(\rho,\rho_0,t) = \frac{1}{2\pi\sqrt{4\pi Dt}} {\rm e}^{-Dt} 
     \frac{\sinh{\frac{\rho_0\rho}{2\pi Dt}}}{\sinh\rho_0\sinh\rho}
     \exp \Bigl[ -\frac{\rho^2+\rho_0^2}{4Dt} \Bigr].   \label{eq.dif5}
 \end{eqnarray}

From Eq.(\ref{eq.dif5}), the following equation is obtained;
\begin{eqnarray}
  f(\rho,t) = \lim_{\rho_0 \rightarrow 0}f(\rho,\rho_0,t)  
            = \bigl( 4\pi Dt \bigr)^{-3/2} {\rm e}^{-Dt}
     \frac{\rho}{\sinh\rho}
     \exp \Bigl[ -\frac{\rho^2}{4Dt} \Bigr].   \label{eq.dif6}
 \end{eqnarray}

In Eq.~(\ref{eq.dif5}), $\rho_0$ denotes the radial rapidity of an emission center.  
It would be identified to the radial flow rapidity. We have estimated radial flow rapidity using Eq.~(\ref{eq.dif5})~\cite{suzu05}.  

If the distribution of $\rho_0$ is taken into account, it would be reasonable to assume that it distributes randomly with dispersion $\sigma_0^2$;
\begin{eqnarray*}
  f(\rho_0,t) = ( 2\pi \sigma_0^2)^{-3/2} {\rm e}^{-\sigma_0^2/2} \frac{\rho_0}{\sinh\rho_0}
     \exp \Bigl[ -\frac{\rho_0^2}{2\sigma_0^2} \Bigr].   
 \end{eqnarray*}
Then the distribution function of radial rapidity is given by the following equation,
\begin{eqnarray}
  f_c(\rho,t) &=& \int_0^\infty f(\rho,\rho_0,t)f(\rho_0,t)\sinh^2 \rho_0 \, \sin\theta d\rho_0 d\theta d\phi \nonumber \\
    &=& ( 2\pi \sigma_T^2)^{-3/2} {\rm e}^{-\sigma_T^2/2} \frac{\rho}{\sinh\rho}
        \exp \Bigl[ -\frac{\rho^2}{2\sigma_T^2} \Bigr],  \label{eq.dif7} \\   
   \sigma_T^2 &=& 2Dt + \sigma_0^2.                      \label{eq.dif8}
 \end{eqnarray}
Equation (\ref{eq.dif7}) is obtained form Eq.~(\ref{eq.dif6}), if $2Dt$ is replaced by $\sigma_T^2$.
Parameter $\sigma_T^2$ is connected to the moment of $\rho_0$ as,
\begin{eqnarray*}
    \langle \rho_0^2 \rangle = \int_0^\infty \rho_0^2 f(\rho_0,t)
                  \sinh^2 \rho_0 \, \sin\theta d\rho_0 d\theta d\phi 
                = \sigma_0^4 + 3 \sigma_0^2.       \label{eq.dif9}
\end{eqnarray*}

When $\rho<<1$, $|{\bf p}|=m\sinh\rho\simeq m\rho$. Then, Eq.~(\ref{eq.dif6}) reduces to
 \begin{eqnarray}
  f(\rho,t)\simeq \exp[-{\rho^2}/(2\sigma(t)^2) ], \quad \sigma(t)^2=2Dt.  
 \label{eq.par1}
 \end{eqnarray}
If we assume that Eq.~(\ref{eq.par1}) should coincide with the Maxwell-Boltzmann distribution, 
we have an identity, $ k_BT=m\sigma(t)^2$, where $k_B$ is the Boltzmann constant. 
Then, Eq.~(\ref{eq.dif8}) reduces to 
 \begin{eqnarray}
  \sigma_T^2 =  k_BT/m +\Bigl(- \frac{3}{2}+\frac{1}{2}\sqrt{9+4\langle \rho_0^2\rangle} \Bigr).   \label{eq.par3}
 \end{eqnarray}
When, $\sigma_0^2<<1$ or $\rho_0^2<<1$, Eq.~(\ref{eq.dif8}) is written as, 
\begin{eqnarray}
   \sigma_T^2 \simeq  k_BT/m + \langle\rho_0^2\rangle/3.       \label{eq.par4}
 \end{eqnarray}
\section{Analysis of large $p_T$ distributions of charged particles}

Transverse momentum distributions of charged particles observed by STAR~\cite{adam03} and PHENIX~\cite{adle04}  Collaborations are analyzed by Eq.~(\ref{eq.dif7}).
Data are taken at $\theta=\pi/2$. In this case, the identity, $\rho=y_T=\xi$, is satisfied.

\begin{figure}[h]
 \begin{minipage}{0.45\linewidth}
    \includegraphics[scale=0.45,bb=50 200 530 630,clip]{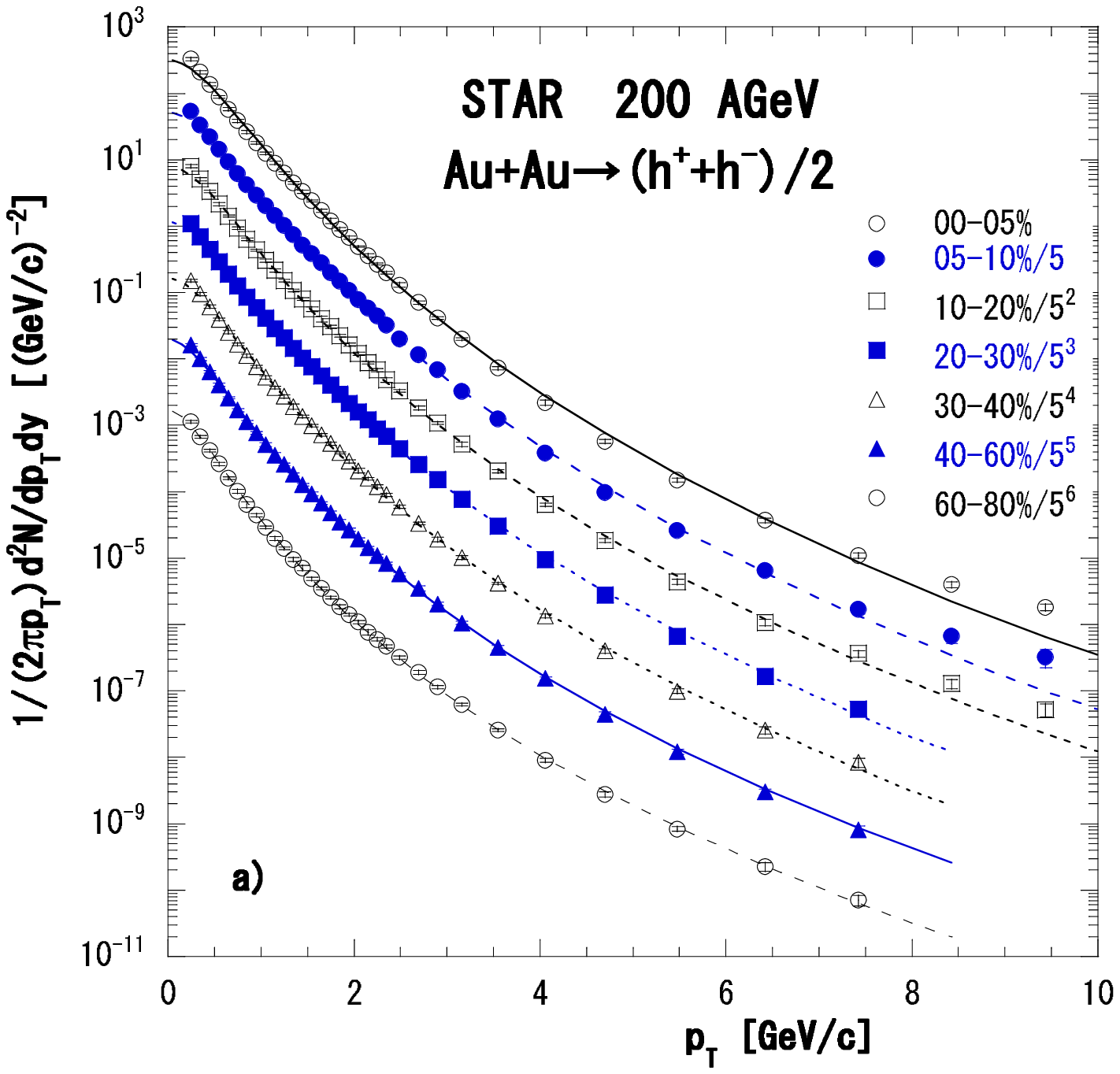}
 \end{minipage}
 \begin{minipage}{0.45\linewidth}
    \includegraphics[scale=0.45,bb=32 210 530 650,clip]{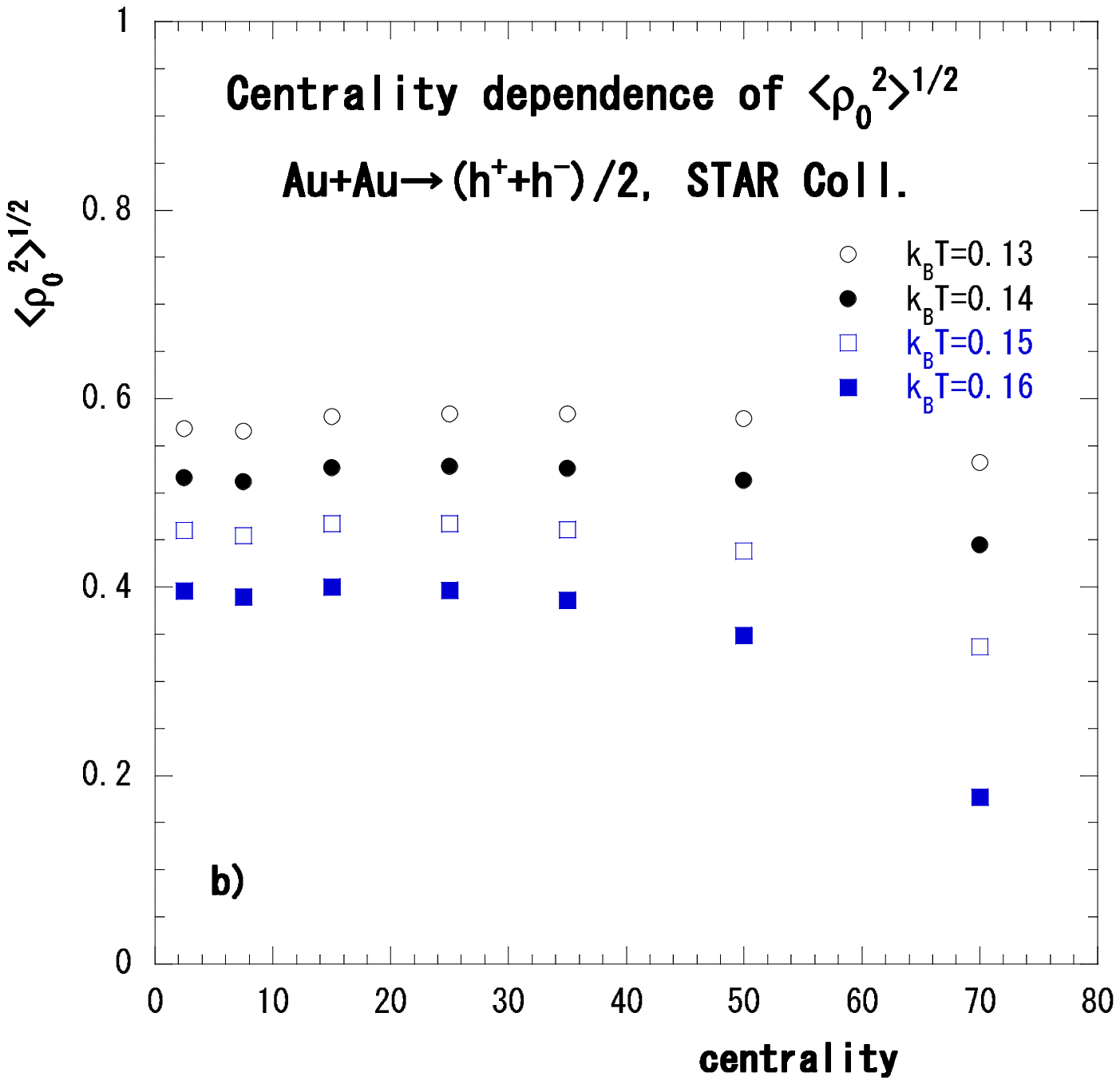}
    \caption{(a) $p_T$ distribution for $Au+Au\rightarrow (h^++h^-)/2+X$ at $y=0$~\cite{adam03}, and (b) centrality dependence of $\langle \rho_0^2 \rangle^{1/2}$  extracted from $p_T$ distributions }
 \label{fig.stach}
 \end{minipage}
\end{figure}

The results on the data by STAR Collaboration are shown in Fig.~\ref{fig.stach} and Table~\ref{tab.table1}.
%
%
\begin{table}[htb]
 \begin{tabular}{cccccc} 
   \hline
    \tablehead{1}{c}{b}{centrality}
  & \tablehead{1}{c}{b}{$C$}
  & \tablehead{1}{c}{b}{$\sigma_T^2$}
  & \tablehead{1}{c}{b}{$m$}
  & \tablehead{1}{c}{b}{$\chi^2_{min}/$n.d.f}\\
   \hline
  00-05\%  & 1126.3  $\pm$ 51.0   & 0.332$\pm$ 0.003 & 0.570$\pm$ 0.007 & 244.9/32\\ 
  05-10\%  & 916.2   $\pm$ 47.5   & 0.332$\pm$ 0.004 & 0.567$\pm$ 0.009 & 128.8/32\\ 
  10-20\%  & 780.0   $\pm$ 36.2   & 0.351$\pm$ 0.004 & 0.536$\pm$ 0.008 & 144.2/32\\ 
  20-30\%  & 590.1   $\pm$ 28.9   & 0.360$\pm$ 0.004 & 0.519$\pm$ 0.008 & 98.1/32\\ 
  30-40\%  & 451.5   $\pm$ 22.7   & 0.374$\pm$ 0.004 & 0.492$\pm$ 0.008 & 95.4/32\\ 
  40-60\%  & 300.99  $\pm$ 14.47  & 0.402$\pm$ 0.004 & 0.442$\pm$ 0.006 & 51.7/32\\ 
  60-80\%  & 150.91  $\pm$  7.43  & 0.444$\pm$ 0.004 & 0.369$\pm$ 0.006 & 26.5/32\\ 
  pp(NSD)  & 17.654  $\pm$ 1.424  & 0.455$\pm$ 0.006 & 0.324$\pm$ 0.008 & 16.1/29\\ \hline
 \end{tabular}
 \caption{Parameters on $p_T$ distributions estimated by Eq.~(\ref{eq.dif7})  
    in $Au+Au\rightarrow (h^++h^-)/2+X$ at $y=0$ at $\sqrt{s_{NN}}=200$ GeV~\cite{adam03}}
 \label{tab.table1} 
\end{table}
%
Information on radial flow is extracted by the use of  Eq.~(\ref{eq.par3}) under the assumption that the hadronization temperature is constant irrespective of centrality. 
The results are shown in Fig.1b.  
The mean radial flow rapidity, $\langle \rho_o^2 \rangle^{1/2}$, depends on centrality very 
weakly, if it is less than 60\%. At $k_BT=0.13$ GeV, $\langle \rho_o^2 \rangle^{1/2}\simeq 0.57$ 
for centrality < 60\%.

\begin{figure}[h]
 \begin{minipage}{0.45\linewidth}
    \includegraphics[scale=0.45,bb=50 200 530 630,clip]{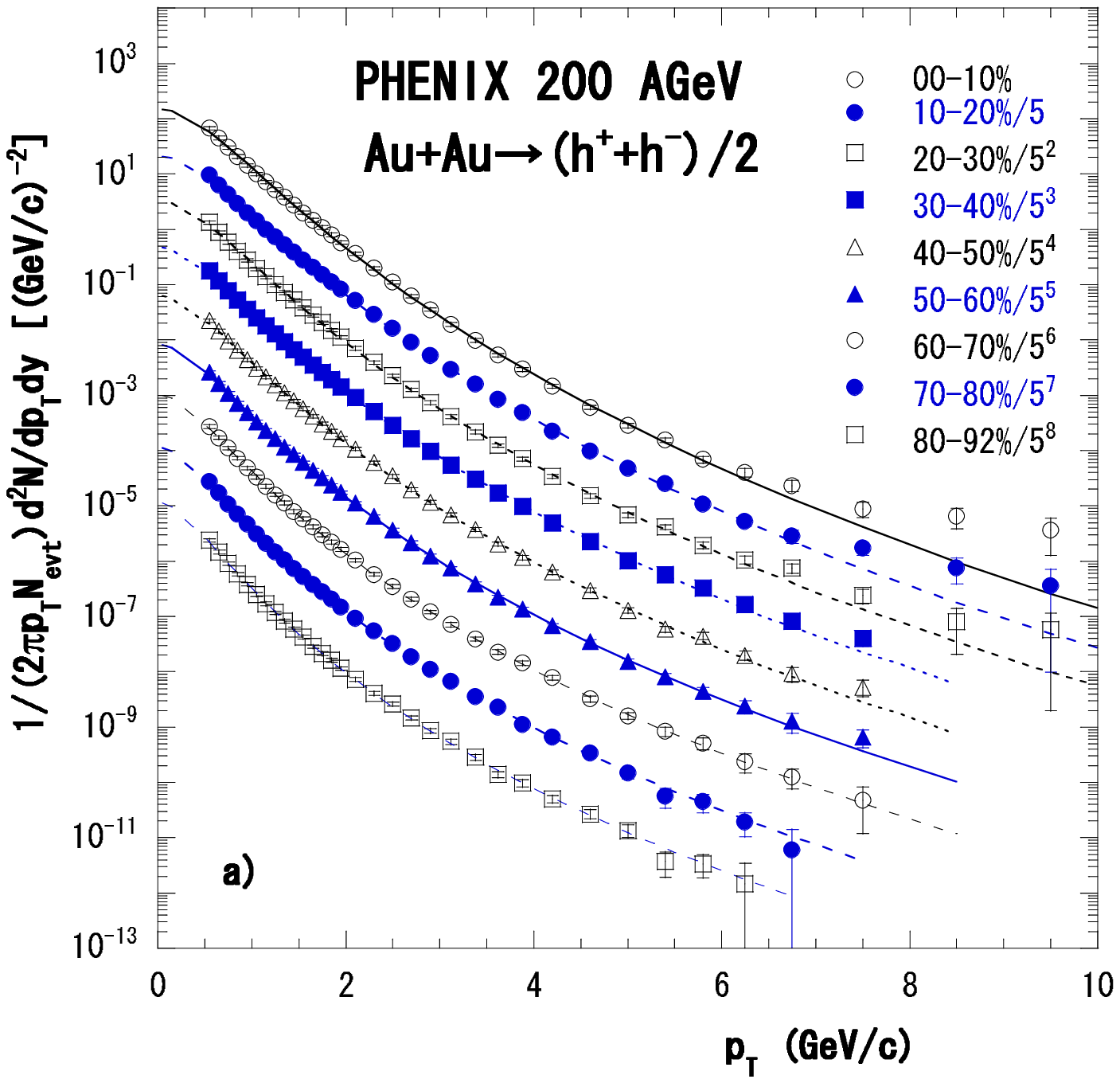}
 \end{minipage}  
 \begin{minipage}{0.45\linewidth}
    \includegraphics[scale=0.45,bb=32 210 530 650,clip]{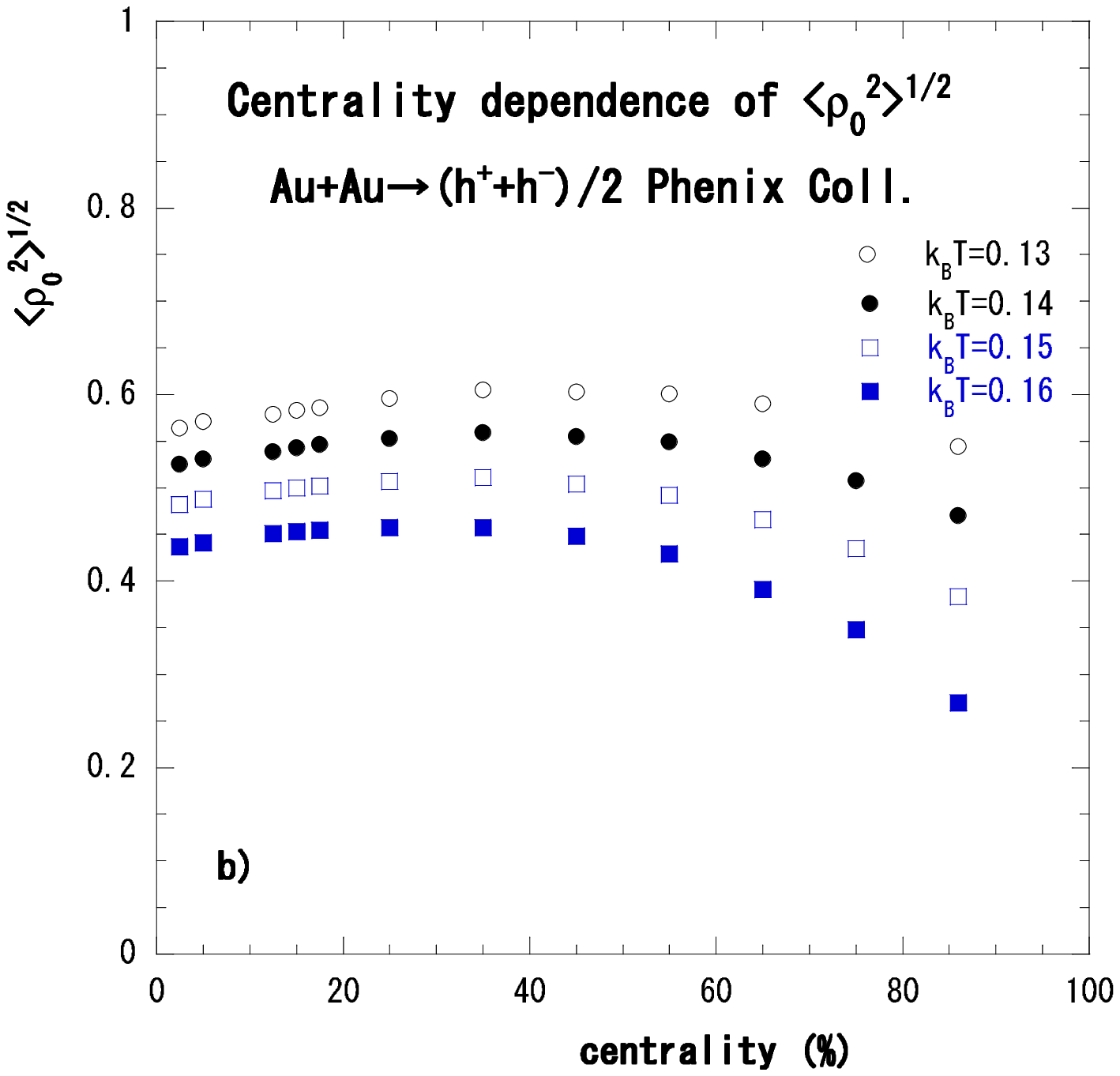}
    \caption{(a) $p_T$ distributions for $Au+Au\rightarrow (h^++h^-)/2+X$ at $y=0$~\cite{adle04}, 
    and (b) centrality dependence of $\langle \rho_0^2 \rangle^{1/2}$  
    extracted from $p_T$ distributions }
      \label{fig.phech}
 \end{minipage}
\end{figure}

The results on the data by FHENIX Collaboration are shown in Fig.~\ref{fig.phech} and 
Table~\ref{tab.table2}. Estimated value of $\langle \rho_o^2 \rangle^{1/2}$ is somewhat larger 
than that from STAR Collaboration at the same temperature.
%
%
%
\begin{table}[htb]
 \begin{tabular}{cccccc} 
  \hline
      \tablehead{1}{r}{b}{centrality}
  & \tablehead{1}{c}{b}{$C$}
  & \tablehead{1}{c}{b}{$\sigma_T^2$}
  & \tablehead{1}{c}{b}{$m$}
  & \tablehead{1}{c}{b}{$\chi^2_{min}/$n.d.f}\\
   \hline 
  00-05\%  & 410.00  $\pm$ 42.99  & 0.277$\pm$ 0.007 & 0.745$\pm$ 0.023 & 32.1/31\\ 
  00-10\%  & 408.50  $\pm$ 41.25  & 0.284$\pm$ 0.007 & 0.726$\pm$ 0.021 & 35.5/31\\ 
  10-20\%  & 305.21  $\pm$ 30.84  & 0.293$\pm$ 0.007 & 0.708$\pm$ 0.021 & 26.4/31\\ 
  20-30\%  & 265.03  $\pm$ 27.18  & 0.313$\pm$ 0.007 & 0.653$\pm$ 0.019 & 27.8/31\\ 
  30-40\%  & 209.37  $\pm$ 21.82  & 0.330$\pm$ 0.007 & 0.611$\pm$ 0.019 & 16.8/29\\ 
  40-50\%  & 143.61  $\pm$ 15.42  & 0.338$\pm$ 0.007 & 0.587$\pm$ 0.018 & 15.4/29\\ 
  50-60\%  & 103.03  $\pm$ 11.65  & 0.358$\pm$ 0.008 & 0.537$\pm$ 0.018 & 9.9/29\\ 
  60-70\%  & 70.343  $\pm$ 8.419  & 0.380$\pm$ 0.009 & 0.485$\pm$ 0.017 & 5.1/29\\ 
  70-80\%  & 40.688  $\pm$ 5.284  & 0.390$\pm$ 0.010 & 0.457$\pm$ 0.018 & 7.4/28\\ 
  80-92\%  & 22.286  $\pm$ 3.126  & 0.406$\pm$ 0.011 & 0.419$\pm$ 0.018 & 10.3/27\\ \hline
 \end{tabular}
 \caption{ Parameters on $p_T$ distributions estimated by Eq.~(\ref{eq.dif7})  
    in $Au+Au\rightarrow (h^++h^-)/2+X$ at $y=0$ at $\sqrt{s_{NN}}=200$ GeV~\cite{adle04}}
\label{tab.table2}
\end{table}
%

\section{High transverse momentum limit of the model}
In the 1970's, when large $p_T$ distributions are observed in accelerator experiments, many models, 
which have power-law behavior in $p_T$ are proposed. In Ref.~\cite{hage83}, a model for $p_T$ distribution, inspired by the QCD, is proposed as $(p_0/(p_T+p_0))^n$.
It approaches an exponential distribution of $p_T$ for $p_T \rightarrow 0$, and a power function of $p_T$ for $p_T \rightarrow \infty$.

In this section, high transverse momentum limit of Eq.~(\ref{eq.dif7}) is examined 
at $\theta=\pi/2$, where the identity $\rho=\ln((m_T+p_T)/m)$ holds.
Radial rapidity contained in the Gaussian-like part in Eq.~(\ref{eq.dif7}) is rewritten as,
 \begin{eqnarray}
    \exp \Bigl[ -\frac{\rho^2}{2\sigma_T^2} \Bigr] = 
        \Bigl(\frac{m_T + p_T}{m}\Bigr)^{-\rho/{(2\sigma_T^2)}}.  \label{eq.hig1}  
 \end{eqnarray}
For any positive number $\epsilon$, we have, 
\begin{eqnarray*}
   \lim_{p_T\rightarrow\infty} \frac{\rho}{p_T^\epsilon}
    = \lim_{p_T\rightarrow\infty} \frac{1}{p_T^\epsilon} \ln\Bigl(\frac{m_T+p_T}{m}\Bigr) = 0.
 \end{eqnarray*}
 Therefore, we can approximate $\rho$ in the exponent in Eq.~(\ref{eq.hig1}) as a constant, which is written by $2c_0$, within some finite transverse momentum range.  Then Eq.~(\ref{eq.dif7}) reduces to, 
\begin{eqnarray}
  f_c(\rho,t) \sim \Bigl(\frac{m}{m_T+p_T}\Bigr)^{c_0/{\sigma_T^2}}\frac{m}{p_T} 
      \sim \Bigl(\frac{m}{2p_T}\Bigr)^{c_0/{\sigma_T^2}+1}. \label{eq.hig2}
 \end{eqnarray}
From Eq.~(\ref{eq.hig2}), one can see that Eq.~(\ref{eq.dif7}) shows the power-law behavior 
in the high transverse momentum limit, and that the power becomes smaller as $\sigma_T^2$,  
which should increase with the colliding energy $\sqrt{s_{NN}}$, increases.

\section{Summary and discussions}

In order to analyze large $p_T$ distributions of charged particles observed at RHIC, the relativistic stochastic process in the three dimensional rapidity space, which is non-Euclidean, 
is introduced. The solution is Gaussian-like in radial rapidity, where the radial flow rapidity $\rho_0$ is included.  It is very similar to the formula proposed in Ref.~\cite{minh73} 
at $\theta=\pi/2$. ( See also Ref.~\cite{biya05}. )

Transverse momentum distributions for charged particles in $Au+Au$ collisions at $y=0$ 
at $\sqrt{s_{NN}}=200$ GeV are analyzed.
From the observed $p_T$ distributions, an effect of radial flow is subtracted by the use of assumption that the hadronization temperature does not depend on the centrality.

The averaged radial flow rapidity, $\langle \rho_0^2 \rangle^{1/2}$, depends on the centrality very weakly from 0\% up to 60\%. It decreases rapidly, as the centrality increases above 60\%.

We have derived that our formula (\ref{eq.dif7}) shows power-law behavior in $p_T$ in the high transverse momentum limit. Therefore, it behaves like the Gaussian distribution in $p_T$ when $p_T<<m$, and like the power-law distribution 
in $p_T$ when $p_T>>1$. 

\begin{theacknowledgments}
 Authors would like to thank RCNP at Osaka University, Faculty of Science, Shinshu University,
  and Matsumoto University for financial support.
\end{theacknowledgments}


\bibliographystyle{aipproc}   

\begin{thebibliography}{9}
%
\bibitem{minh73} M.~Duong-van and P.~Carruthers, \emph{Phys. Rev. Letters}, \textbf{31}, 133 (1973) 
%
\bibitem{suzu04}  N.~Suzuki and M.~Biyajima, \emph{Acta Phys. Polon.} \textbf{B35}, 283 (2004); hep-ph/0404112 
%
\bibitem{adam03} J.~Adams, et al., STAR Collaboration, \emph{Phys. Rev. Letters}, \textbf{91}, 172301 (2003)
%
\bibitem{adle04} S.S.~Adler, et al., PHENIX Collaboration, \emph{Phys. Rev.}, \textbf{C69}, 034910 (2004)
%
\bibitem{suzu04b} N.~Suzuki and M.~Biyajima, math-ph/0406040, unpublished
%
\bibitem{suzu05} N.~Suzuki and M.~Biyajima, hep-ph/0504076 
%
\bibitem{hage83} R.~Hagedorn,  \emph{CERN preprint} Ref.TH.3684-CERN (1983)
%
%
\bibitem{biya05}M.~Biyajima, M.~Kaneyama, T.~Mizoguchi and G.~Wilk, \emph{Euro Phys. J.}, \textbf{C40}, 243 (2005) %
\end{thebibliography}

\end{document}